\documentclass{elsart}
\usepackage{amsmath}
\usepackage{amssymb}
\usepackage[english]{babel}

\newcommand{\set}[1]{\left\{#1\right\}}
\newcommand{\iset}[1]{\mathcal{#1}}
\newcommand{\pr}[1]{\left(#1\right)}
\newcommand{\spr}[1]{\left[#1\right]}
\newcommand{\abs}[1]{{\left|#1\right|}}

\newcommand{\enset}[2]{\left\{#1 ,\ldots , #2\right\}}
\newcommand{\vect}[1]{\spr{#1}}

\newcommand{\np}{\textbf{NP}}
\newcommand{\pp}{\textbf{PP}}
\newcommand{\fnp}{\textbf{FNP}}
\newcommand{\fpnp}{\textbf{FP$^\text{NP}$}}
\newcommand{\entr}[1]{\mathcal{E}\pr{#1}}

\newcommand{\threequart}{\frac{3}{4}}

\newtheorem{prb}{Problem}

\newenvironment{problem}[1]{\begin{prb}\textup{\textbf{(#1)}}}{\end{prb}}

\begin{document}
\begin{frontmatter}

\title{Computational Complexity of Queries Based on Itemsets}
\author{Nikolaj Tatti}
\address{HIIT Basic Research Unit,
Laboratory of Computer and Information Science,
Helsinki University of Technology, Finland}
\maketitle
\begin{abstract}
We investigate determining the exact bounds of the frequencies of conjunctions based on frequent sets. Our scenario is an important special case of some general probabilistic logic problems that are known to be intractable. We show that despite the limitations our problems are also intractable, namely, we show that checking whether the maximal consistent frequency of a query is larger than a given threshold is \np-complete and that evaluating the Maximum Entropy estimate of a query is \pp-hard. We also prove that checking consistency is \np-complete.
\end{abstract}
\end{frontmatter}
\section{Introduction}
Assume that we have two events, say $a$ and $b$. Assume further that their probabilities are $P(a) = 0.6$ and $P(b) = 0.5$. What can we say about the probability of $a \land b$? We know that the probability must lie within $I = \spr{0.1, 0.5}$. This interval is tight: For each $f \in I$ there is a distribution having $f$ as a probability of $a \land b$. Also note that the Maximum Entropy estimate in this case is $0.6\times0.5 = 0.3$.

A more complicated example would be the following: Assume three events $a_1$, $a_2$, and $a_3$. Assume that we know $P(a_1)$, $P(a_2)$, $P(a_3)$, $P(a_1 \land a_2)$ and $P(a_1 \land a_3)$. What can we say about $P(a_1 \land a_2 \land a_3)$?

Let us make these examples more general: A conjunctive query is a boolean formula having the form $a_1 \land a_2 \land \ldots \land a_L$. Assume that we are given a set $\iset{F}$ of conjunctive queries along with their probabilities. Assume also that we are given a conjunctive query $B$ not belonging to $\iset{F}$. What can we tell about the probability of this query? We know that the possible probabilities of the query $B$ correspond to some interval. In the paper we show that checking whether the right side of this interval is larger than some threshold is \np-complete. We also show that estimating the probability of the query $B$ using Maximum Entropy is \pp-hard.

In the paper we adopt the terminology used in data mining of $0$--$1$ data: Conjunctive queries are represented by sets of items called itemsets and the probabilities of conjunctive queries are called itemset frequencies.

Our problems are special cases of much more general problems (see Section~\ref{sec:freqsat} for detailed comparison). These general problems are well-studied and they are all (at least) \np-hard. The difference is that in our work we concentrate on studying antimonotonic families of itemsets.  We should point out that antimonotonic families are important since they tend to arise frequently in practice, for example, in mining of frequent itemsets~\cite{agrawal93mining,agrawal96apriori}. A similar technique is used in~\cite{cooper90complexity} to prove that inference of Belief Networks is \np-hard. The result of~\cite{cooper90complexity} is essentially Theorem~\ref{thr:entropy} (in this paper) though it is in a different context. The general boolean query scenario is reduced to Linear Programming in~\cite{hailperin65inequalities}. A method worth mentioning is introduced in~\cite{pavlov03beyond} where the authors estimate the frequencies using Maximum Entropy.


\section{Preliminaries}
In this section we give basic definitions used in mining of $0$--$1$ data.

By a \emph{binary data set} we mean a collection of binary vectors of length $K$ sampled from some distribution. We define a \emph{sample space} $\Omega = \set{0,1}^K$ to be the collection of all possible binary vectors of length $K$. From now on $\Omega$ will always denote the sample space, $K$ will denote the dimension of binary vectors. Any distribution given in this paper will be defined on $\Omega$.

It is custom to assign an attribute to each dimension of $\Omega$. Thus, when we speak of $a_i$ we mean the $i$th dimension. The set of all attributes is $A = \enset{a_1}{a_K}$. An \emph{itemset} is a subset of $A$. Let $B = \enset{a_{i_1}}{a_{i_L}}$ be an itemset. We often use a condensed notation $B = a_{i_1}\cdots a_{i_L}$. A family of itemsets is called \emph{antimonotonic} if all the subsets of any member are also included.

Let $p$ be a distribution defined on $\Omega$. We use the following notation: Let $B = a_{i_1}\cdots a_{i_L}$ be an itemset and let $t$ be a binary vector of length $L$. Then we shorten the notation $p(a_{i_1} = t_1,\ldots , a_{i_L} = t_L)$ by $p(B = t)$. By $p(B = 1)$ we mean $p(B = t)$, where $t$ contains only ones. The probability $p(B = 1)$ is called the \emph{frequency} of $B$.

Assume a family $\enset{B_1}{B_N}$ of itemsets and a vector $\theta$ of length $N$. We say that a distribution $p$ \emph{satisfies} the frequencies if $\theta_i = p(B_i = 1)$ for $i = 1,\ldots,N$. We say that these frequencies are \emph{consistent} if there is a distribution satisfying them.

\section{Maximal Frequency Query is NP-complete}
Assume that we want to find the frequency for an itemset $B$ based on some known family $\iset{F}$ of itemsets. We know that generally the frequency for $B$ is not unique: There may be distributions that produce different frequencies for $B$ but have the same frequencies of $\iset{F}$. The set of all the consistent frequencies of $B$ is an interval~\cite{bykowski03support}. In this section we focus on finding one side of this interval:

\begin{problem}{MaxQuery}
Assume that we are given an antimonotonic family $\iset{F}$ having $N$ members along with rational and consistent frequencies $\theta$. Find the maximal frequency for a given itemset $B$ that can be produced by a distribution satisfying the frequencies $\theta$.
\end{problem}
In other words, we ask ourselves that, if we know the frequencies $\theta$, then what is the largest consistent frequency for $B$. Note that the maximal frequency always exists since the frequencies $\theta$ are required to be consistent. Our goal in this section is to show that in general this problem is intractable. First let us give an example where the solution can be easily obtained.
\begin{exmp}
Assume that a family $\iset{F}$ contains only the itemsets of size one. Then the frequency $\theta_{a_i}$ is the mean of the attribute $a_i$. The maximal frequency for an itemset $B = b_1b_2\cdots b_M$ is $\min\set{\theta_{b_i} \mid i = 1,\ldots,M}$.
\end{exmp}

We know that \textsc{MaxQuery} can be solved by using Linear Programming~\cite{bykowski03support} though the resulting program contains an exponential number of variables. This reduction along with some results from Linear Programming theory~\cite{papadimitriou98combi} has important consequences: There is a distribution, say $q$, producing the maximal frequency for B and having at most $N+1$ non-zero entries. Also, $q$ has rational entries, and if $L$ is the number of bits needed to specify the denominator of an element of the frequency vector $\theta$, then the number of bits needed to specify the denominator of an entry of $q$ is $\log_2\pr{(N+1)^32^{NL}} \in O(NL)$. We call such a distribution \emph{canonical}.

Since {\np} is defined for yes/no problems we need the decision version of \textsc{MaxQuery}:
\begin{problem}{MaxQueryDec}
Assume that we are given an antimonotonic family $\iset{F}$ having $N$ members along with rational and consistent frequencies $\theta$. Given an itemset $B$ and a rational threshold $b$ is there a distribution satisfying the frequencies $\theta$ such that the frequency of $B$ is larger than $b$?
\end{problem}
The relation between \textsc{MaxQuery} and \textsc{MaxQueryDec} is the following: Assume that we can solve \textsc{MaxQuery} in polynomial time, then we can clearly solve \textsc{MaxQueryDec} in polynomial time. Assume now that we can solve \textsc{MaxQueryDec} in polynomial time. Let $f$ be the solution of \textsc{MaxQuery}. We can find $f$ using \textsc{MaxQueryDec} and dichotomous search. We know that $f$ is a rational number between $0$ and $1$ and that the denominator of $f$ can be expressed using $O(NL)$ bits. Thus the number of required search steps is $O(NL)$.

\begin{thm}
\textsc{MaxQueryDec} is in \np.
\label{thr:maxquerynp}
\end{thm}
\begin{pf}
Let $q$ be a canonical distribution for \textsc{MaxQuery}. We can represent this distribution in polynomial space, and hence we can use it as a certificate. To check the certificate we need to check that $q$ is a real distribution, that it satisfies the frequencies and that its frequency for $B$ is larger than the threshold $b$.
\end{pf}
Our next step is to reduce \textsc{3SAT} to \textsc{MaxQueryDec}. In order to do that we need the following lemma:
\begin{lem}
Assume that two distributions $p$ and $q$ satisfy the frequencies $\theta$ of an antimonotonic family $\iset{F}$ of itemsets. Let $C \in \iset{F}$. Then $p(C = t) = q(C = t)$ for any binary vector $t$.
\label{lem:projected}
\end{lem}
\begin{pf} Fix $C = \enset{c_1}{c_N}$ and $t$. Let $U = \set{c_i \in C \mid t_i = 1}$ and let $W = C - U$. Denote the elements of $W$ by $w_i$. Let $p(U = 1, \bigvee_i w_i = 1)$ be the probability of $U$ being $1$ and at least one of $w_i$ being $1$. We see that 
\begin{equation}
p(C = t) = p(U = 1, W = 0) = p(U = 1) - p(U = 1, \bigvee w_i = 1).
\label{eq:lem1eq1}
\end{equation}
Let $\mathcal{H} = \set{H \subseteq W \mid H \neq \emptyset}$ be the collection of non-empty subsets of $W$. We can express the last term of Eq.~\ref{eq:lem1eq1} by using the inclusion-exclusion principle
\begin{equation}
p(U = 1, \bigvee w_i = 1) = \sum_{H \in \mathcal{H}} (-1)^{\abs{H}+1}p(U = 1, H = 1).
\label{eq:lem1eq2}
\end{equation}
By combining Eqs.~\ref{eq:lem1eq1}~and~\ref{eq:lem1eq2} we have expressed $p(C = t)$ as a linear combination of terms having the form $p(B = 1)$ where $B \subseteq C$. Antimonotonicity implies that all these frequencies are included in $\theta$. This makes $p(C = t)$ unique and the lemma follows.
\end{pf}
\begin{thm}
\textsc{3SAT} is polynomial-time reducible to \textsc{MaxQueryDec}.
\label{thr:maxquerycomplete}
\end{thm}
\begin{pf}
Let $R$ be an instance of \textsc{3SAT} having $L$ variables and $M$ clauses. We set the dimension of the sample space to be $K = L+M$. The first $L$ items correspond to the variables of $R$ and the last $M$ items correspond to the clauses. We use the following notation: Let $t$ be a truth assignment and let $C_i$ be a clause, then $C_i(t)$ is a function resulting $1$, if $C_i$ is satisfied by $t$, and $0$ otherwise. We denote the first $L$ items by $v_i$ and the last $M$ items by $c_i$. We also set $V = \enset{v_1}{v_L}$ and $W = \enset{c_1}{c_M}$.

We will now define an antimonotonic family $\iset{F}$ of itemsets. Let $C_i$ be some clause and let $c_i$ be its corresponding item. Assume that the items corresponding to the variables in $C_i$ are $v_1$, $v_2$, and $v_3$. We add an itemset $v_1v_2v_3c_i$ to the family $\iset{F}$ along with its subsets. We repeat this procedure to each clause in $R$. The resulting family $\iset{F}$ contains $16M$ members at maximum.

The following step is to define the frequencies $\theta$. In order to do this we define a distribution $p$ over the attributes to be
\[
p(V = t, W = u) =
\left\{\begin{array}{ll}
2^{-L} & \text{if for all } i \text{ we have } u_i = C_i(t)\\
0 & \text{otherwise}.
\end{array}\right.
\]
That is, the first $L$ items are distributed uniformly and the values of the last $M$ items are set to correspond to the truth values of the clauses.

We define the frequencies $\theta_i = p(F_i = 1)$, where $F_i \in \iset{F}$. We note that the frequencies are rational and consistent. There is a closed formula for evaluating these frequencies. For example, assume that we have a clause $C_1  \equiv (v_1 \lor v_2 \lor v_3)$. The frequency of the itemset $v_1v_2v_3c_1$ is then
\[
\sum_{t,u}p(V = t,W = u) = \sum_{t,u_i = C_i(t)}p(V = t, W = u) = 2^{L-3}2^{-L} = \frac{1}{8},
\]
where in the first summation $t$ ranges over truth assignments such that $t_1 = t_2 = t_3 = 1$ and $u$ ranges over binary vectors of length $M$ such that $u_1 = 1$. In the second summation $t$ ranges similarly as in the first summation and $u$ is now set to correspond to the clauses. The frequencies for the other members of $\iset{F}$ can be deduced in a similar way. Thus we can obtain the frequencies $\theta$ in polynomial time.

Let $f$ be the maximal frequency for the itemset $W$. We claim that the formula $R$ is satisfiable if and only if $f > 0$.

Assume that $R$ is satisfiable by a truth assignment, then we have
\[
f = p(W = 1) \geq p(V = t, W = 1) = 2^{-L} > 0.
\]

Assume now that there is a distribution $q$ satisfying the frequencies and producing a positive frequency for $W$. Let $t$ be a truth assignment not satisfying the formula, that is, there is a clause, say $C_1 = (v_1 \lor v_2 \lor v_3)$, that is not satisfied. Define $G = v_1v_2v_3$ and $u = \vect{t_1, t_2, t_3}$. Lemma~\ref{lem:projected} implies that $q(V = t, W = 1) \leq q(G = u, c_1 = 1) = p(G = u, c_1 = 1)= 0$. By reversing this property we get the following: If $t$ is such that
\begin{equation}
q(V = t, W = 1) > 0
\label{eq:satisfy}
\end{equation}
holds, then $t$ must satisfy $R$.

By the assumption $q(W = 1) > 0$ so there exists a truth assignment $t$ such that Eq.~\ref{eq:satisfy} holds. Thus $R$ is satisfiable. The reduction is complete if we set the query $B = W$ and the threshold $b = 0$.
\end{pf}
\begin{exmp}
Consider the formula $(v_1 \lor v_2) \land (\neg v_2 \lor v_3)$. We have two clauses, $C_1$ and $C_2$, and three variables, $v_1$, $v_2$, and $v_3$. The itemset family along with its frequencies (given in parenthesises) is
\[
\iset{F} = \left\{
\begin{array}{l}
\emptyset\pr{1}, v_1\pr{\half}, v_2\pr{\half}, v_3\pr{\half}, v_1v_2\pr{\quart}, v_2v_3\pr{\quart}, \\
c_1\pr{\threequart}, v_1c_1\pr{\half}, v_2c_1\pr{\half}, v_1v_2c_1\pr{\quart}, \\
c_2\pr{\threequart}, v_2c_2\pr{\quart}, v_3c_2\pr{\half}, v_2v_3c_2\pr{\quart}
\end{array}
\right\}.
\]
The maximal frequency of $c_1c_2$ for this setup (solved by linear programming) is $\half$. Clearly, the formula is satisfiable.
\end{exmp}

\section{MaxEnt Frequency Query is PP-hard}
In the previous section we showed that searching for the maximal frequencies is a very hard problem. The maximal frequencies, however, are not so useful if our goal is to estimate boolean queries from a given set of itemsets. A much more useful approach is to use Maximum Entropy approach. Given a distribution $p$ defined on $\Omega$, the \emph{entropy} of $p$ is $\entr{p} = -\sum_{\omega \in \Omega} p(\omega)\log\pr{p(\omega)}$. It is custom to define $0\log(0) = 0$ so that $\entr{p}$ is always defined. 
\begin{problem}{EntrQuery}
Assume that we are given an antimonotonic family $\iset{F}$ having $N$ members along with rational and consistent frequencies $\theta$. Find a frequency for a given itemset $B$ produced by the distribution $p$ satisfying the frequencies $\theta$ and maximising the entropy $\entr{p}$.
\end{problem}
It has been empirically shown that \textsc{EntrQuery} results in a good approximation~\cite{pavlov03beyond}.

Again we need a decision version of the problem:
\begin{problem}{EntrQueryDec}
Assume that we are given an antimonotonic family $\iset{F}$ having $N$ members along with rational and consistent frequencies $\theta$. Let $f$ be a frequency for a given itemset $B$ produced by a distribution satisfying the frequencies $\theta$ and maximising entropy. Is $f$ larger than a given rational threshold $b$?
\end{problem}

The following theorem shows that \textsc{EntrQueryDec} is \np-hard.
\begin{thm}
\textsc{3SAT} is polynomial-time reducible to \textsc{EntrQueryDec}.
\label{thr:entropy}
\end{thm}
\begin{pf}
Let $R$ be an instance of \textsc{3SAT}. Let $\iset{F}$, $\theta$, $V$ and $B$ be the same as in the proof of Theorem~\ref{thr:maxquerycomplete}. Let $\mathbb{P}$ be the set of distributions satisfying the frequencies $\theta$. Let $q \in \mathbb{P}$. A marginal distribution $q_V$ is obtained from $q$ by keeping only the items included in $V$. The distribution $q$ has the following property: The items corresponding to the clauses are completely determined by the items corresponding to the variables. This implies that the entropy of $\entr{q} = \entr{q_V}$~\cite[Theorem 4.2]{kullback68information}.

Let $\hat{q} \in \mathbb{P}$ be the distribution maximising the entropy. Let $p \in \mathbb{P}$ be the distribution defined in the proof of Theorem~\ref{thr:maxquerycomplete}. Note that $\entr{\hat{q}_V}  = \entr{\hat{q}} \geq \entr{p} = \entr{p_V}$. We know that there is no distribution that has larger entropy than the uniform distribution~\cite[Theorem 3.1]{kullback68information}. Since $p_V$ is uniform, we must have $\entr{\hat{q}_V} = \entr{p_V}$. Hence $\entr{\hat{q}} = \entr{p}$. We also know that the distribution maximising entropy is unique~\cite[Theorem 3.1]{csiszar75divergence}. This implies that $\hat{q} = p$. To complete the proof we note that $p$ produces a positive frequency for $B$ if and only if $R$ is satisfiable.
\end{pf}
A problem \textsc{P} is in {\pp} if there is a machine such that an input $x$ is a yes-instance of \textsc{P} iff more than half of the computation paths end up accepting~\cite{papadimitriou95complexity}. The class {\pp} is (believed to be) larger than \np. We can show that \textsc{EntrQueryDec} is \pp-hard: In the proof the frequency of $B$ is exactly the number of satisfying assignments divided by $2^{-L}$. Hence, if we set the threshold $b = 2^{-L/2}$, the instance will be in \textsc{EntrQueryDec} iff the square root of the number of assignments satisfy the given \textsc{3SAT} formula. This problem is known to be \pp-complete~\cite{bailey01phase}.

\section{Checking Consistency is NP-complete}
So far we have assumed that the itemset frequencies given in our problems are consistent. Let us remove this constraint and consider the following problem.
\begin{problem}{Consistent}
Assume that we are given an antimonotonic family $\iset{F}$ having $N$ members along with rational frequencies $\theta$. Are the frequencies $\theta$ consistent?
\end{problem}

The following theorem proves that \textsc{Consistent} is a very hard problem.
\begin{thm}
\textsc{Consistent} is \np-complete.
\end{thm}
\begin{pf}
First, we need to show that \textsc{Consistent} is in \np. We know from Linear Programming theory that if the frequencies are valid then there is a canonical distribution satisfying the frequencies. This is our certificate and thus \textsc{Consistent} is in \np.

We now prove that \textsc{3SAT} is polynomial-time reducible to \textsc{Consistent}. We use the same construction as in the proof of Theorem~\ref{thr:maxquerycomplete} with some additions: We add one special attribute, say $c_0$, to the set of attributes. We add an itemset $c_0$ to $\iset{F}$, and we also add itemsets having the form $c_0c_i$ to $\iset{F}$. The frequencies for the new itemsets are set to be $2^{-L}$, where $L$ is the number of variables appearing in the \textsc{3SAT} instance $R$.

Assume that $R$ is satisfiable by a truth assignment $t$. We define a distribution $q$ by extending the distribution $p$ to $c_0$. The extension is done such that $c_0$ is $1$ iff $V = t$. Clearly, $q$ satisfies the frequencies.

To prove the other direction, assume that there exists a distribution, say $q$, that satisfies the frequencies. To prove that $R$ is satisfiable we must prove that $q(W = 1) > 0$. Select two attributes, say $c_1$ and $c_2$. Note that $q(c_0 = 1, c_1 = 0) = 0$ and $q(c_0 = 1, c_2 = 0) = 0$. This implies that $q(c_0 = 1) = q(c_0 = 1, c_1 = 1, c_2 = 1)$. We can prove in an iterative fashion that
\[
q(W = 1) \geq  q(c_0 = 1, W = 1) = q(c_0 = 1) = 2^{-L}.
\]
This proves the result.
\end{pf}
\section{Connections to Related Work}
\label{sec:freqsat}
An \np-complete problem called \textsc{FreqSat} introduced in~\cite{calders03thesis,calders04computational} is a generalisation of \textsc{Consistent} --- in \textsc{FreqSat} we are allowed to have non-antimonotonic families and inequality constraints. We can transform \textsc{MaxQueryDec} into \textsc{FreqSat} by changing the query into an inequality constraint. We should also point out that the proof of \np-hardness of \textsc{FreqSat} given in~\cite{calders03thesis} is (although not explicitly mentioned) actually a valid proof for \textsc{Consistent}.

An even more general scenario is introduced in~\cite{lukasiewicz01logic} in which we are allowed to have conditional first-order logic sentences as constraints/queries. This scenario can be emulated by itemsets~\cite{calders04computational}. Also, a famous problem called \textsc{PSat} in which we are given a CNF-formula, a frequency for each clause, and we are asked whether there is a distribution satisfying the frequencies is known to be \np-complete~\cite{georgakopoulos88psat}.
\section{Conclusions}
In this paper we studied certain boolean query problems. Our problems were specialised (but frequently occurring and thus important) problems of much general scenarios and we showed that despite the limitations our problems remained intractable. The crux of the paper lies within the construction in the proof of Theorem~\ref{thr:maxquerycomplete}.

There are some open problems: For example, what is the exact complexity of \textsc{MaxQuery}? Is it \fnp-complete or \fpnp-complete? Also, what is the complexity of the opposite problem \textsc{MinQuery}? In addition, it is worthwhile to study the conditions under which the boolean query problems can be solved efficiently.

\bibliographystyle{plain}
\bibliography{bibliography}
\end{document}